\documentclass{jfm}

\usepackage{graphicx}
\usepackage{float}

\usepackage{subfigure}
\usepackage{color}
\usepackage{psfrag}
\usepackage{natbib}
\usepackage{multirow}
\usepackage{psfrag}
\usepackage{xfrac}
\usepackage{sidecap}
\usepackage{booktabs}

\usepackage{enumitem}
\usepackage{pstricks}
\usepackage{tikz}
\usepackage{amsmath}

\usepackage{calc}

\usepackage{caption}
\usepackage{rotating}
\usepackage{multirow}
\usepackage{tabularx}
\usepackage{upgreek}
\usepackage[normalem]{ulem}
 \usepackage{siunitx}

\usepackage{bm}
\usepackage{mathrsfs}
\usepackage{bigdelim}
\usepackage{pifont}
\usepackage{txfonts} 
\usepackage{afterpage}
\usepackage{lscape}
\def\Rey{\mbox{Re}}   
\def\Sh{\mbox{Sh}}   
\def\Nu{\mbox{Nu}}   
\def\Sc{\mbox{Sc}}   %
\def\Pr{\mbox{Pr}}   %
\def\RH{\mbox{RH}}   

\begin{document}

\author{Mogeng Li$^1$, Detlef Lohse$^{1,2}$, and Sander G. Huisman$^1$}

\title{\color{black} High humidity enhances the evaporation of non-aqueous volatile sprays}

\shorttitle{High humidity enhances the evaporation of non-aqueous volatile sprays}
\shortauthor{M. Li, S. G. Huisman and D. Lohse}

\affiliation{\aff{1}Physics of Fluids Group, Max Planck Center for Complex Fluid Dynamics, J.\,M.\,Burgers Center for Fluid Dynamics, Department of Science and Technology, University of Twente, 7500AE Enschede, The Netherlands
\aff{2}Max Planck Institute for Dynamics and Self-Organisation, 37077 G{\"o}ttingen, Germany}

\date{Received: date / Accepted: date}

\maketitle

\begin{abstract}

We experimentally investigate the evaporation of very volatile liquid droplets (Novec 7000 Engineered Fluid) in a turbulent spray. Droplets with diameters of the order of a few micrometers are produced by a spray nozzle and then injected into a purpose-built enclosed dodecahedral chamber, where the ambient temperature and relative humidity in the chamber are carefully controlled. We observe water condensation on the rapidly evaporating droplet, both for the spray and for a single acoustically levitated millimetric Novec 7000 droplet. We further examine the effect of humidity, and reveal that a more humid environment leads to faster evaporation of the volatile liquid, as well as more water condensation. This is explained by the much larger latent heat of water. We extend an analytical model based on Fick's law to quantitatively account for the data.

\end{abstract}


\section{Introduction} \label{sec:Intro}

Evaporation of volatile sprays plays an important role in many practical applications, such as fuel injection for combustion and spray cooling. For a single droplet in a quiescent environment, assuming that the evaporation process is limited by the diffusion of the vapour layer, the droplet diameter can be described by the so-called $d^2$-law \citep{langmuir1918evaporation}. The evaporation rate depends on many flow parameters and material properties, including the ambient temperature, vapour pressure of the liquid, as well as the vapour concentration in the ambient environment. The evaporating droplet cools down due to its latent heat. Its temperature asymptotes to a constant value lower than the ambient temperature, which can be evaluated by balancing the absorption of latent heat in the liquid-to-gas phase transition with the heat transfer between the droplet and the surrounding air \citep{dalla2021revisiting}. When a stationary droplet is exposed to an external turbulent flow, the turbulent convection will accelerate the transport of vapour away from the droplet, resulting in larger concentration gradients and thus in higher evaporation rates and shorter droplet life times \citep{birouk2006current}. On the other hand, when the droplet itself is moved by the turbulence, it may pass through regions with different vapour concentrations along its trajectory, and the instantaneous droplet evaporation rate is affected accordingly.

When a multicomponent droplet is considered, a few complications are introduced alongside. Different components evaporating at different rates, creating a concentration gradient in the liquid phase. The evaporation process is affected not only by the vapour diffusion rate in air as in the single-component scenario, but also by the interspecies diffusion of different vapours, multicomponent phase equilibrium relations (typically Raoult's law), and the diffusion and convection in the liquid phase \citep{sirignano2010fluid}. The vapourisation of multicomponent droplets has been studied extensively in the context of combustion, especially in the scenario of a mixture of various fuels vapourising at an elevated temperature \citep{daif1998comparison, ra2009vaporization, promvongsa2017multicomponent}.

For a non-aqueous single-component droplet evaporating in a humid environment at room temperature, it has been observed that water vapour in the ambient air condenses on the evaporating droplet, forming a multicomponent droplet \citep{law1987alcohol, marie2014lagrangian}, provided that the evaporative cooling process can lead to a droplet temperature low enough to saturate the water vapour. The size evolution of the multicomponent droplet in these studies was measured by direct imaging and digital holography, respectively, and described by an analytical model, where the two mass transfer equations are coupled through a heat transfer equation. 

In this study, we further demonstrate that the ambient humidity affects the evaporation of a non-aqueous volatile droplet through the condensation process. The experimentally measured water condensation droplet size is compared with model predictions, and the effect of turbulent convection is discussed with an analytical model.

The paper is organised as follows: we first provide details of the experimental method in \S\ref{sec:Exp_setup}. Measurements of droplet sizes in a spray and the analytical model are covered in \S\ref{sec:results}. The effect of ambient turbulence is further discussed in \S\ref{sec:turb} and finally the observation of a single millimetric droplet evaporating in an acoustic levitator is detailed in \S\ref{sec:single}. The paper ends with conclusions and an outlook.

\section{Experiment setup} \label{sec:Exp_setup}
\begin{figure}
    \centering
\setlength{\unitlength}{1cm}
\begin{picture}(13.5,8.5)
\put(0,0){\includegraphics[width = 0.55\textwidth]{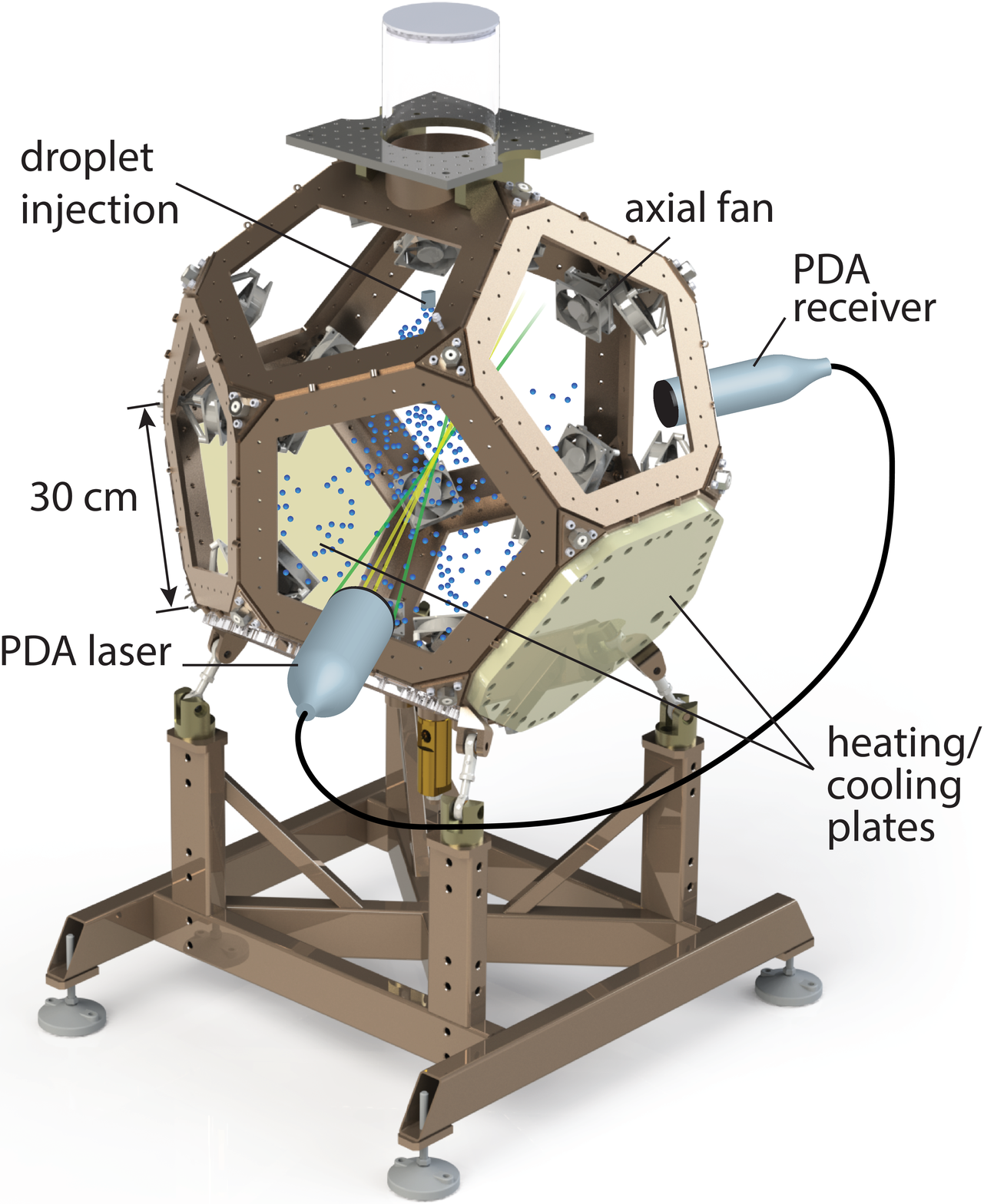}}
\put(8,4.8){\includegraphics[width = 0.35\textwidth]{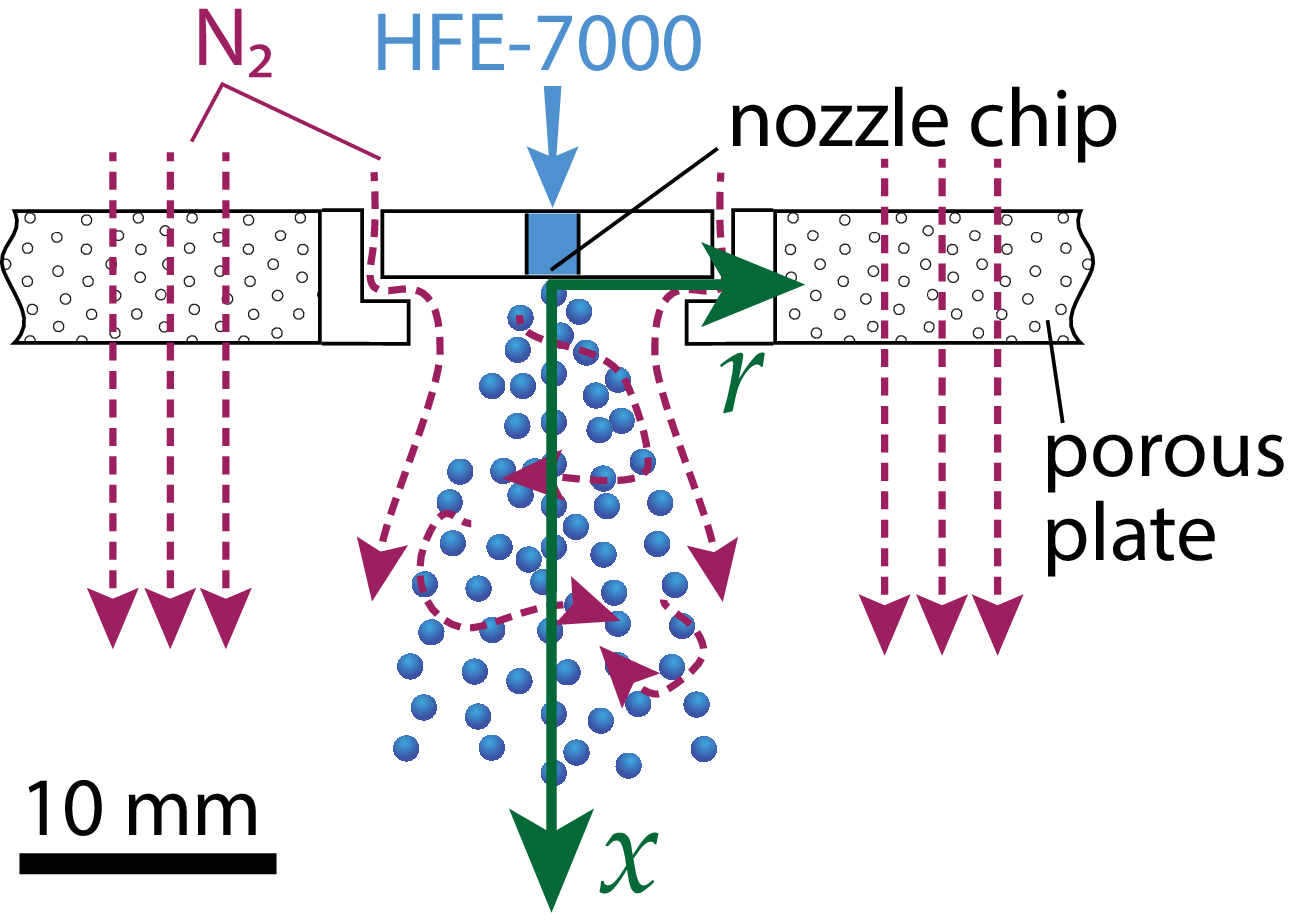}}
\put(8,1){\includegraphics[width = 0.35\textwidth]{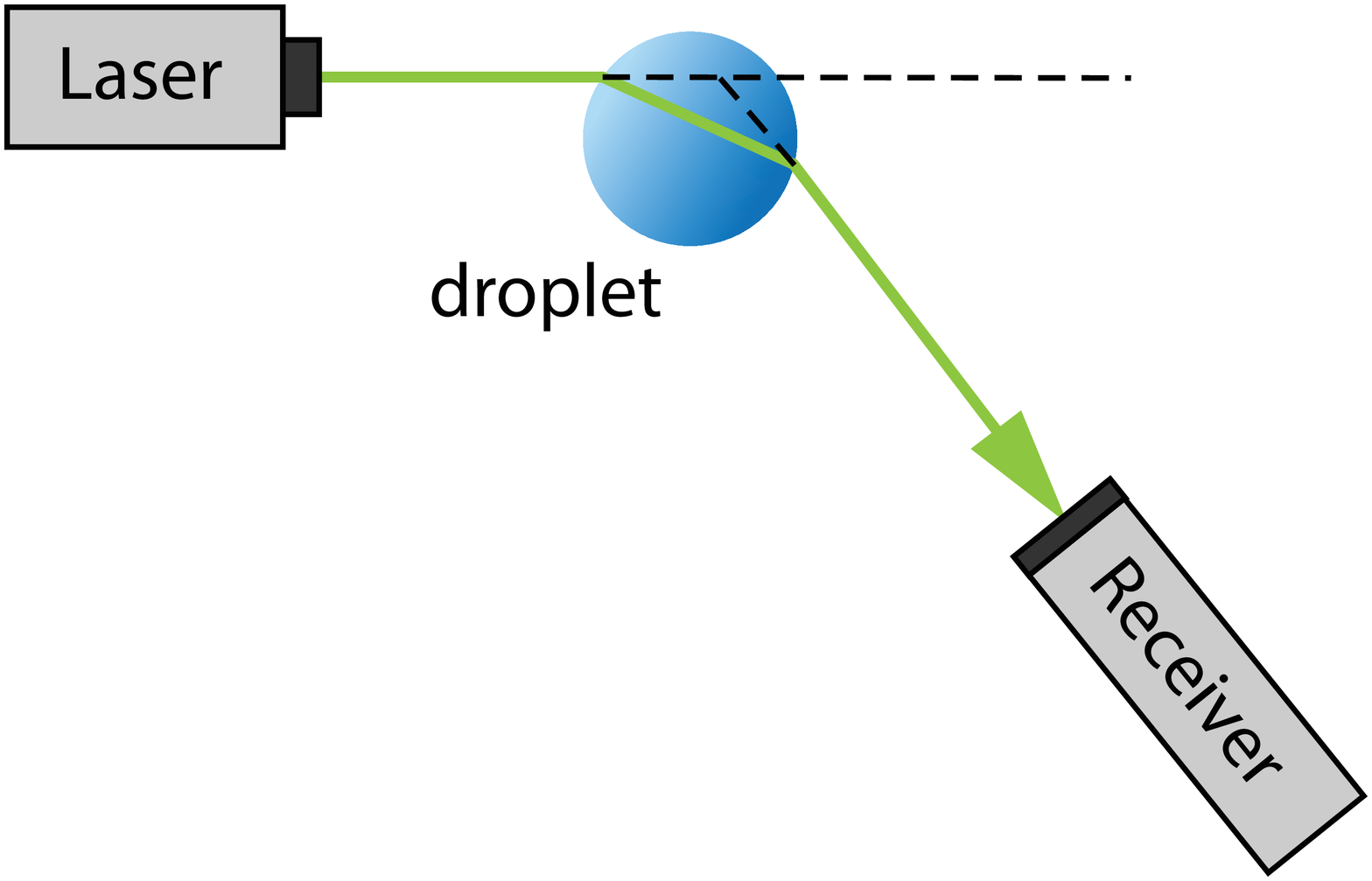}}
\put(1.5,7.5){(\textit{a})}
\put(7.3,7.5){(\textit{b})}
\put(7.3,3.8){(\textit{c})}
\put(11,3.4){$\theta \approx 60^{\circ}$}
\end{picture}
    \caption{(\textit{a}) The dodecahedral chamber with PDA measurement system. The chamber is enclosed by 12 pentagonal transparent acrylic window panels (not shown in the drawing). The droplets are injected through a nozzle chip from the top of the chamber. (\textit{b}) Schematic showing a close-up view of the nozzle provided by Medspray. The nozzle chip contains 90 holes with a diameter of $4.5$ $\mu$m. (\textit{c}) Laser optical path, with the first order refraction mode captured by the receiver. The droplets in all panels are not shown to scale. }
    \label{fig:setup}
\end{figure}

A dodecahedral chamber has been constructed to investigate this problem. A schematic of the setup is shown in figure \ref{fig:setup}(\textit{a}). The chamber has 12 pentagonal faces with a side length of $30~$cm, resulting in a volume of $\approx210~$L. To generate a homogeneous isotropic turbulent flow field in the centre of the dodecahedron, 20 identical axial fans are mounted at the vertices and driven by the same DC power. All fans are rigged to simultaneously blow into the centre of the dodecahedron. The temperature in the chamber can be regulated through the two heating/cooling plates on the side, which are connected to an external heater/cooler Julabo FP51-SL capable of achieving a temperature range of $-30\,\text{--}\,80^{\circ}$C. The relative humidity (\RH) can also be controlled within a range between $20\%$ and $100\%$, by either injecting water using an air humidifier or removing excess moisture by pumping the air through an in-line dryer filled with desiccants. To avoid any disturbance to the flow field, \RH{} is adjusted beforehand and it remains relatively unchanged during each measurement thanks to the enclosed nature of the chamber. The temperature, \RH, and atmospheric pressure are sampled using a BME280 sensor with an in-house built Arduino data acquisition system.

The nozzle used to generate the spray is provided by Medspray. As illustrated in figure \ref{fig:setup}(\textit{b}), droplets are formed at the nozzle chip driven by a syringe pump, and carried downstream by a $15~$L/min co-flow to prevent coalescence. We choose to use Novec 7000 Engineered Fluid (chemical name hydrofluoroethers HFE-7000, hereinafter referred to as `HFE-7000' in short) manufactured by 3M \citep{NovecTDS}, and also characterised by \cite{rausch2015density}. In essence, HFE-7000 is extremely volatile, with a very high saturation vapour pressure of $52.7~$kPa at $20^{\circ}$C, and it is relatively harmless to the ozone layer and the acrylic windows of the dodecahedron.

The droplet diameters are measured using a Dantec two-component Phase Doppler Anemometry (PDA) system with a 112 mm FiberFlow probe. Schematics of the setup and the optical arrangement are shown in figures \ref{fig:setup}(\textit{a}) and (\textit{c}). The receiving probe is positioned at an off-axis angle of $\theta\approx60^{\circ}$, where the first order refraction is the dominant mode. The measurement volume is fixed at the centre of the dodecahedron, while the spray nozzle is traversed to obtain axial and radial profiles. At each measurement location, the PDA signal is sampled for $30~$s. The concentration of HFE-7000 vapour in the ambient environment is considered negligible, as during each run less than $2~$mL of the liquid is injected, whereas $650~$mL is needed to saturate the entire volume of the dodecahedron.

\section{Experimental results on droplet evaporation in a spray} \label{sec:results}
\begin{figure}
    \centering
\setlength{\unitlength}{1cm}
\begin{picture}(12.5,5)
\put(0,0){\includegraphics[scale = 0.9]{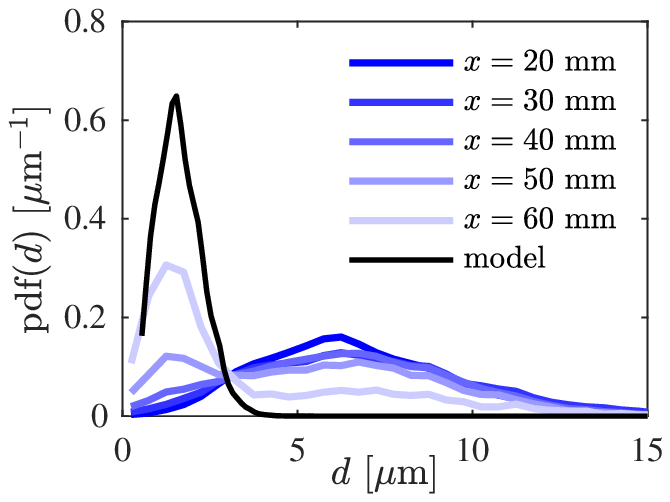}}
\put(6.75,0){\includegraphics[scale = 0.9]{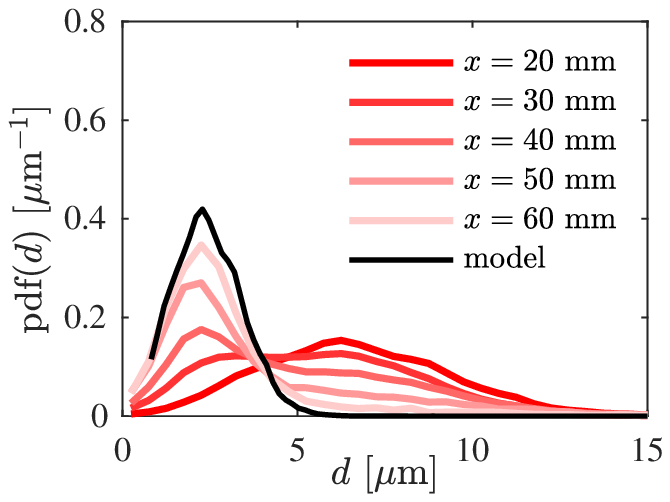}}
\put(0,4.7){(\textit{a})}
\put(2.7,4.5){$\RH = 20\%$}
\put(6.75,4.7){(\textit{b})}
\put(9.45,4.5){$\RH = 100\%$}
\put(4,1.5){\vector(-1,-1){0.7}}
\put(3.5,1.6){increasing $x$}
\put(10.75,1.5){\vector(-1,-1){0.7}}
\put(10.25,1.6){increasing $x$}
\end{picture}
    \caption{Droplet size distribution at (\textit{a}) $\RH = 20\%$ and (\textit{b}) $\RH = 100\%$. From dark to light, the line colour indicates an increasing distance $x$ from the nozzle along the centreline, ranging from $x = 20$ mm to $x = $60 mm. The solid black line is the predicted condensed water droplet size distribution based on an analytical model (see Appendix \S \ref{sec:app}), valid for large distances $x$.}
    \label{fig:pdf}
\end{figure}

We first explore the change in the average droplet diameter along the centreline ($x$ coordinate) of the spray jet. Figures \ref{fig:pdf}(\textit{a}) and (\textit{b}) show the probability density distributions (pdf) of the measured droplet diameters for the ambient relative humidity cases of $\RH = 20\%$ and $\RH = 100\%$, respectively. Close to the nozzle, the distribution exhibits a broad peak centred around $d=6~\upmu$m. Further away from the nozzle, as shown by the lines with a lighter shade, the magnitude of this peak reduces while a second peak emerges at a much smaller size, $d = 2\mbox{--}3~\upmu$m. This peak at small diameters keeps growing with increasing $x$, eventually exceeding the magnitude of the first peak. We believe that the emergence of the second peak at much smaller diameters is a result of water in the ambient air condensing on the rapidly evaporating HFE-7000 droplets. An analytical model is extended based on the work of \cite{marie2014lagrangian} and \cite{tonini2015novel}, taking the immiscibility between HFE-7000 and water into consideration. Details of the model can be found in Appendix \S \ref{sec:app}. Assuming that the droplets measured at $x = 20~$mm contain HFE-7000 only, the final water droplet size distribution can be predicted, and is shown by the black lines. Indeed, the measured size distribution can be well captured by the model prediction. Note that at this stage, we consider a single, stationary droplet, and the evaporation-condensation rate is limited by the diffusion of the vapour film around the droplet. Although many effects, including the turbulent convection \citep{birouk2006current, mees2020statistical, dodd2021analysis} and the sheltering in clusters of droplets \citep{villermaux2017fine, sahu2018interaction, dodd2021analysis}, are not taken into consideration, the modelled remaining water droplet size distribution still has a good agreement with the measurements.

\begin{figure}
    \centering
\setlength{\unitlength}{1cm}
\begin{picture}(12.5,5)
\put(0,0){\includegraphics[scale = 0.9]{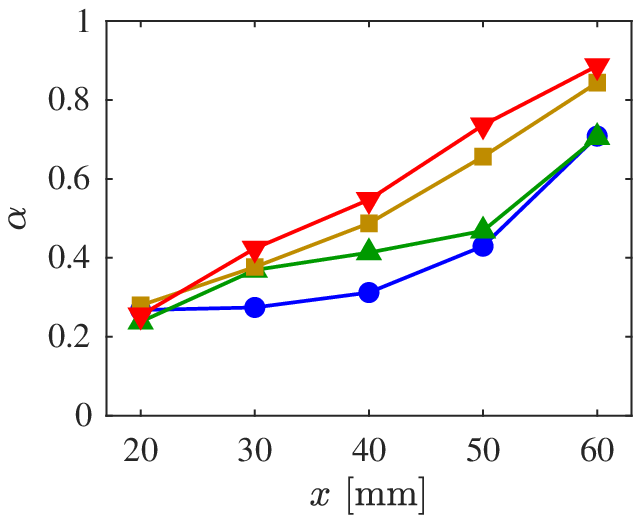}}
\put(6.75,0){\includegraphics[scale = 0.9]{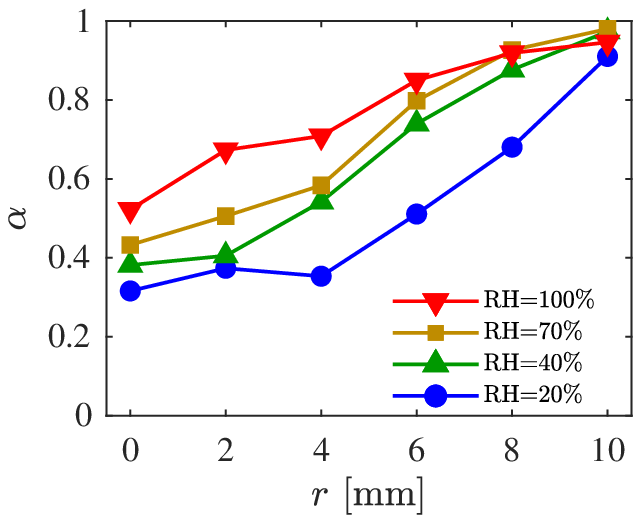}}
\put(0,4.8){(\textit{a})}
\put(6.75,4.8){(\textit{b})}
\end{picture}
    \caption{The fraction $\alpha$ of remaining water droplets among the total number of measured droplets along (\textit{a}) the axial direction ($r = 0$) and (\textit{b}) the radial direction ($x = 40$ mm).}
    \label{fig:alpha}
\end{figure}

\begin{figure}
    \centering
\setlength{\unitlength}{1cm}
\begin{picture}(12.5,7.6)
\put(0,0){\includegraphics[scale = 0.9]{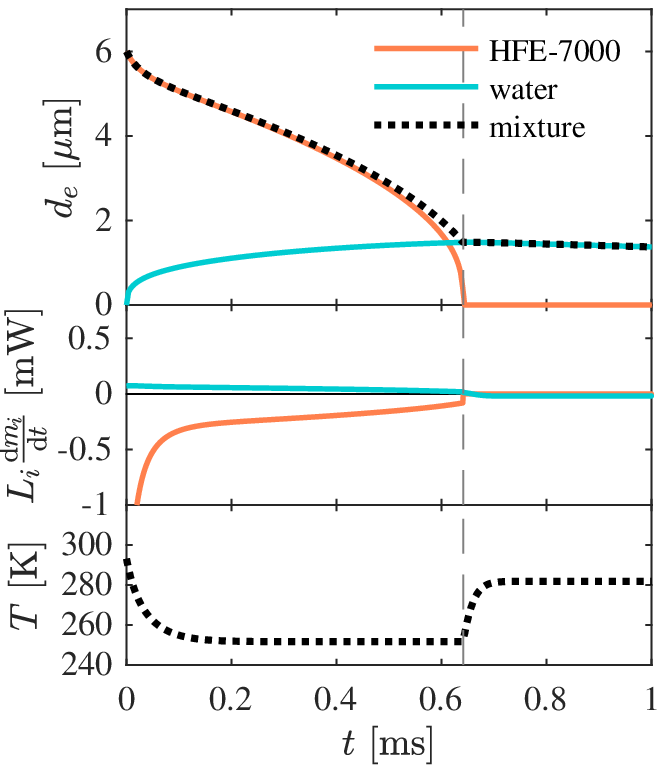}}
\put(6.75,0){\includegraphics[scale = 0.9]{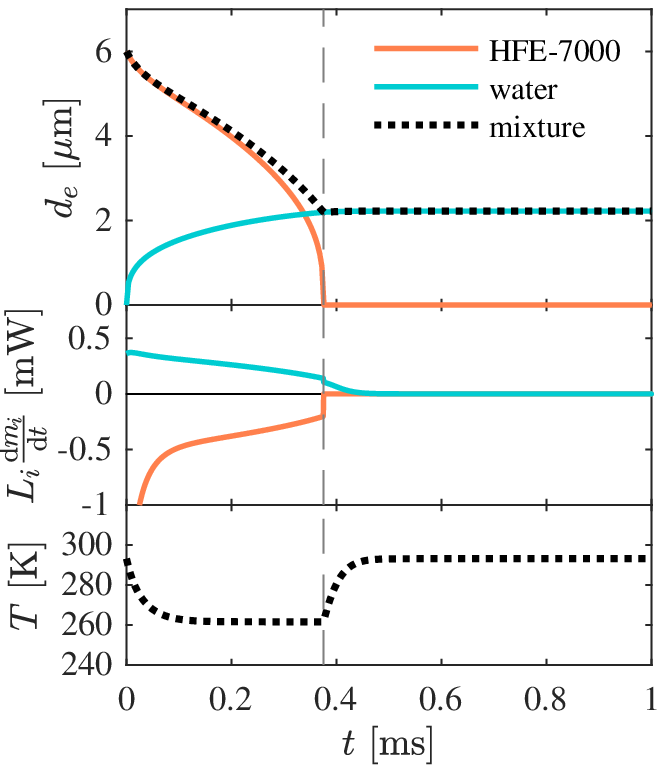}}
\put(1.3,6.4){(\textit{a})}
\put(8.05,6.4){(\textit{b})}
\put(1.3,3.7){(\textit{c})}
\put(8.05,3.7){(\textit{d})}
\put(1.3,1.9){(\textit{e})}
\put(8.05,1.9){(\textit{f})}
\put(2.7,7.1){$\RH = 20\%$}
\put(9.45,7.1){$\RH = 100\%$}
\end{picture}
    \caption{Model prediction of (\textit{a,b}) the equivalent droplet diameter of composing liquids, (\textit{c,d}) the rate of latent heat absorbed/released by the droplet, and (\textit{e,f}) the droplet temperature as a function of time at (\textit{a}, \textit{c}, \textit{e}) $\RH = 20\%$ and (\textit{b}, \textit{d}, \textit{f}) $\RH = 100\%$. The droplet initially contains only HFE-7000 and the ambient temperature is $T_{\infty} = 20^{\circ}\mathrm{C}$ ($293.15$ K), and the initial diameter is set to a typical size of $d_0 = 6~\upmu$m. }
    \label{fig:model_pred}
\end{figure}

We consider the HFE-7000 component fully evaporated in any measured droplet with a diameter smaller than a certain threshold $d_{\text{thr}}$. Thus, a ratio $\alpha$ can be defined as the ratio between the number of droplets with $d<d_{\text{thr}}$ and the total number of droplets measured at the same location. The threshold is chosen as $d_{\text{thr}}=4.7~\upmu$m, which corresponds to the 99.5 percentile of the predicted remaining water droplet diameters in the $\RH = 100\%$ case. Note that the same value for $d_{\text{thr}}$ is used across all \RH{} cases for consistency, although more droplets that still contain liquid HFE-7000 will be mistaken as remaining water droplets in low ambient \RH{} cases. The spatial variation of $\alpha$ at a range of ambient relative humidities is depicted in figure \ref{fig:alpha}. In both axial and radial directions, $\alpha$ increases with increasing distance from the nozzle, confirming the sheltering effect in the jet \citep{villermaux2017fine, wang2021direct}. Furthermore, we observe a strong dependence of $\alpha$ on \RH: at the same location, $\alpha$ increases with \RH, indicating that HFE-7000 evaporates faster in a more humid environment.

The observation can be explained by inspecting the modelled single droplet behaviour. As shown in figures \ref{fig:model_pred}(\textit{a}) and (\textit{b}), indeed, the predicted lifetime of a HFE-7000 droplet at $\RH = 100\%$ is around $40\%$ shorter than that at $\RH = 20\%$. The reason behind this higher evaporation rate lies in the latent heat and the resulting temperature. As shown in figures \ref{fig:model_pred}(\textit{e,f}), when the HFE-7000 droplet is exposed to ambient air, its temperature rapidly decreases as a result of the latent heat absorbed during the liquid-to-gas phase transition. This phenomenon is also known as `evaporative cooling'. The cold droplet cools down its surrounding gas film, reducing the saturation vapour mass fraction of water it carries. As a consequence, the local \RH{} will increase, and once it exceeds 1, the water vapour becomes over-saturated and will start condensing on the surface of the droplet. At a high ambient \RH{} (figure \ref{fig:model_pred}\textit{d}), the rate of water condensation significantly increases and the latent heat released in the process heats up the droplet to a higher temperature, therefore the HFE-7000 component will evaporate at a higher temperature and a faster rate. Furthermore, high \RH{} environment leads to a larger water core, giving rise to a higher surface-to-volume ratio of the HFE-7000 component, which also accelerates the evaporation.

\section{Effect of turbulence} \label{sec:turb}

\begin{figure}
    \centering
\setlength{\unitlength}{1cm}
\begin{picture}(12.5,5.6)
\put(0,0){\includegraphics[scale = 0.9]{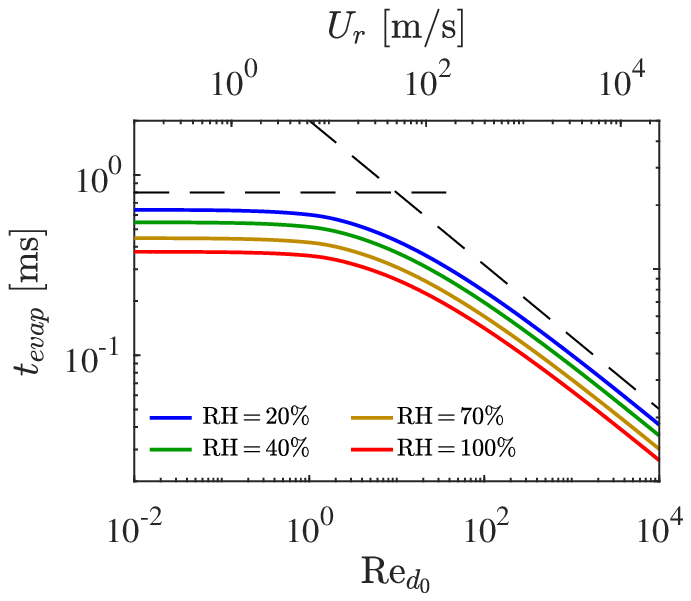}}
\put(6.75,0){\includegraphics[scale = 0.9]{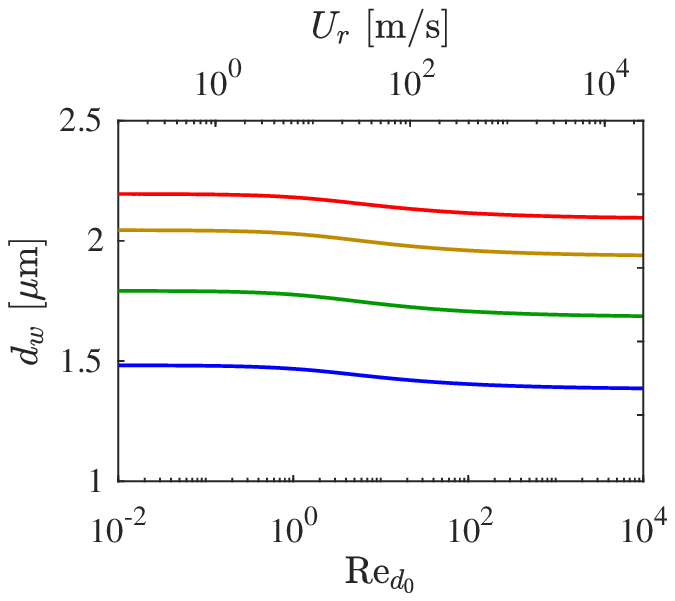}}
\put(0,4.6){(\textit{a})}
\put(6.75,4.6){(\textit{b})}
\put(4.1,3.3){$t_{\text{evap}}\propto \Rey_{d_0}^{-2/5}$}
\put(1.3,3.85){$t_{\text{evap}} = \text{const.}$}
\end{picture}
    \caption{Model prediction at a range of droplet Reynolds numbers, $\Rey_{d_0}$. (\textit{a}) $t_{\text{evap}}$, the time required for the HFE-7000 component to fully evaporate, and (\textit{b}) the diameter of condensed water droplets $d_w$. All results are for a HFE-7000 droplet with an initial diameter of $d_0 = 6~\upmu$m at an ambient temperature of $T_{\infty} = 20^{\circ}$C ($293.15$ K). }
    \label{fig:Re_effect}
\end{figure}

In this section, we discuss the effect of turbulence on the observed evaporation-condensation process. The effect of turbulence is usually modelled through Ranz--Marshall correlations, and is characterised by a single non-dimensional parameter $\Rey_{d}\equiv d U_{r}/\nu$, which is the droplet Reynolds number based on $d$, the diameter, $U_r$, the velocity relative to the surrounding gas, and $\nu$, the kinematic viscosity of air. Since the droplet diameter $d$ changes during evaporation, we represent the level of turbulence using $\Rey_{d_0}\equiv d_0 U_{r}/\nu$, the droplet Reynolds number based on the initial diameter and assume the relative velocity $U_r$ remains constant. In the present experiments, the relative velocity is not measured, nevertheless, $\Rey_{d_0}$ is estimated to be well below 10 considering the small size of the droplets. Figure \ref{fig:Re_effect} compares $t_{\text{evap}}$, the time required for HFE-7000 to fully evaporate and $d_{w}$, the water condensation droplet size at various $\Rey_{d_0}$ values. Overall, $t_{\text{evap}}$ reduces with increasing ambient \RH. For a given \RH, $t_{\text{evap}}$ is also found to be smaller at a higher $\Rey_{d_0}$, as the mass and heat exchange is enhanced by turbulent convection. In the limit of extremely large $\Rey_{d_0}$ which is not practical for the current configuration, a power law relationship $t_{\text{evap}}\propto \Rey_{d_0}^{-2/5}$ emerges as a consequence of (\ref{eq:Sh1}). Interestingly, despite the large variation in $t_{\text{evap}}$, very little change is observed in the final condensed water droplet size $d_w$, as shown in figure \ref{fig:Re_effect}(\textit{b}). This suggests that as turbulent convection enhances the evaporation, the rate of condensation is also increased almost proportionally: similar to its effect in reducing the local HFE-7000 vapour concentration by advecting the HFE-7000 vapour away from the droplet, the turbulence also serves to replenish the droplet surface with over-saturated humid air. This result also explains why even the diffusion-limited model can produce a remaining water droplet size distribution with a high level of agreement with the experimental data.

\begin{figure}
    \centering
\setlength{\unitlength}{1cm}
\begin{picture}(6,5)
\put(0,0){\includegraphics[scale = 0.9]{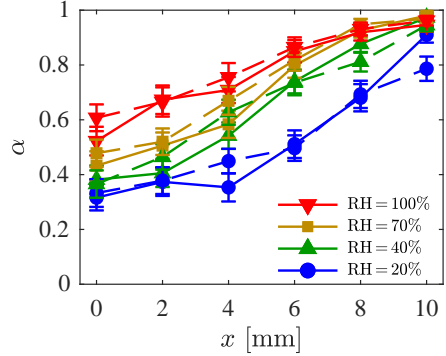}}
\end{picture}
    \caption{The fraction of remaining water droplets among the total number of measured droplets with (dashed line) and without (solid line) the presence of ambient turbulence. The errorbars show the standard deviation of the mean estimated using a bootstrapping approach.}
    \label{fig:turb_exp}
\end{figure}

We would like to point out that the spray is already at a turbulent state even though the surrounding air in the dodecahedral chamber is under quiescent conditions. The ambient turbulence is, in fact, acting on an already turbulent jet spray. Figure \ref{fig:turb_exp} compares $\alpha$, the proportion of the condensed water droplets at a number of measurement locations along the radial direction with quiescent and turbulent ambient conditions. The data with ambient turbulence display a similar trend of increasing $\alpha$ with increasing $x$ and \RH, and there is no major difference from the quiescent case. 

\section{Observations of a single droplet} \label{sec:single}

In this section, we provide further experimental evidence of the water condensation that accompanies the rapid vapourisation process  
by directly imaging a single, millimetre-scale HFE-7000 droplet in an acoustic levitator. The droplet is imaged using a Photron Nova S12 high-speed camera with a Navitar 12$\times$ lens at 500 Hz.

\begin{figure}
    \centering
    \setlength{\unitlength}{1cm}
	\begin{picture}(14,7)
    \put(0,0){\includegraphics[width = 0.95\textwidth]{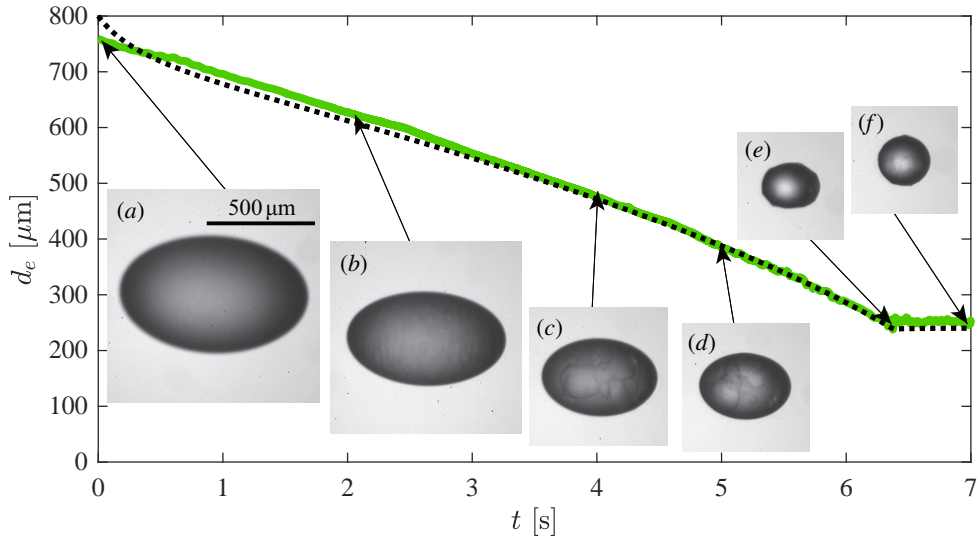}}
    \put(1.45,4.2){(\textit{a})}
    \put(4.4,3.5){(\textit{b})}
    \put(7,2.65){(\textit{c})}
    \put(9,2.45){(\textit{d})}
    \put(9.75,5){(\textit{e})}
    \put(11.25,5.35){(\textit{f})}
    \end{picture}
    \caption{Temporal evolution of the droplet equivalent diameter $d_e$. The solid green line is measured from droplet images, and the dotted black line is computed from the model, with $\Rey_{d_0} = 10$. Insets (\textit{a--f}) are droplet images at time instances marked by the arrows. The scale bar in (\textit{a}) holds for all images.}
    \label{fig:AcousticLevitatorGroup}
\end{figure}

Snapshots of the multi-component droplet are shown in figure \ref{fig:AcousticLevitatorGroup}. The moment the droplet is injected in the acoustic levitator is denoted as $t = 0$. At $t = 2.0~$s (figure \ref{fig:AcousticLevitatorGroup}\textit{b}), small water droplets with a diameter around $20\,\upmu$m distribute within the HFE-7000 drop. Carried by the internal flow, which is caused by the internal circulation as well as the acoustic field, these water droplets move in the HFE-7000 drop, collide, and coalesce into larger clusters as clearly visible in figures \ref{fig:AcousticLevitatorGroup}(\textit{c}) and (\textit{d}). These clusters have complex, non-spherical shapes. When the amount of remaining liquid HFE-7000 is not sufficient to wet the entire surface of the water clusters and fill the grooves, these water clusters are exposed and the drop exhibits an irregular shape with protrusions (figure \ref{fig:AcousticLevitatorGroup}\textit{e}). After this transient phase, only a single-composition water droplet remains (figure \ref{fig:AcousticLevitatorGroup}\textit{f}). It has a spherical shape, as a result of the increase in the surface tension with HFE-7000 being replaced by water at the droplet surface. These observations clearly validate the modelling assumption that the water condensation forms an inclusion within a HFE-7000 shell (see figure \ref{fig:drop_sketch}\textit{b}), and the droplet surface contains predominantly only HFE-7000. Movie 1 shows the entire evaporation-condensation process. Note that for a typical droplet in the spray, the diameter is less than 1\% of the acoustic levitated drop diameter here, and it shape is more spherical even in the multi-component stage because of the much smaller Weber number.

The droplet has an oblate ellipsoidal shape as a result of levitation by the acoustic field \citep{yarin1999evaporation,ALZAITONE2018164}. The major and minor axis lengths $d_a$ and $d_b$ are measured from the image, and the equivalent diameter $d_e$ is calculated as $d_e = \left(d_a^2d_b\right)^{1/3}$. The ratio $d_b/d_a$ is approximately 0.6 until HFE-7000 is fully evaporated. The evaporation time of a spherical droplet with the same equivalent diameter gives a reasonable approximation of the ellipsoidal case \citep{tonini2016one, tonini2019analytical}. The measured diameter variation matches well with the model prediction at $\Rey_{d_0} = 10$, which corresponds to a relative velocity of $U_r \approx 0.2\,$m/s.

\section{Conclusions and outlook} \label{sec:conclusions}

In this work, we experimentally observe an interesting evaporation-condensation process of a volatile spray in humid air. The evaporation of HFE-7000 and the condensation of water can be described by an analytical model based on Fick's law. The immiscibility between HFE-7000 and water is taken into consideration, through a model of the condensation water core contained in the HFE-7000 droplet, as corroborated by the direct imaging of a single droplet in an acoustic levitator. The analytical model also reveals that in a more humid environment, the droplet temperature is higher because more latent heat is released with a higher rate of water condensation under such conditions, confirming the faster HFE-7000 evaporation rate found in experiments. 

In future works, the detailed physics of the water condensation process in the single levitated (or sessile) droplet should be explored: how the condensed water migrates to the interior of the droplet remains an open question. Further, exploring other liquid pairs with high miscibility and different volatilities is also of interest, because they can result in evaporation rates very different from the current findings.

\section*{Acknowledgements}
We acknowledge Wietze Nijdam and Stefan van der Vegte from Medspray for providing the nozzles and technical support during the project. We would also like to thank Uddalok Sen for the single levitated droplet measurements, Gert-Wim Bruggert for the design of the dodecahedron, Dennis van Gils for developing the Arduino data acquisition system, Geert Mentink and Rindert Nauta for building the linear traverse and axial fan power supply, Martin Bos and Thomas Zijlstra for technical support, and Lydia Bourouiba, Andrea Prosperetti, and Chao Sun for discussion.

\section*{Funding}
This work was funded by the Netherlands Organisation for Health Research and Development (ZonMW), project number 10430012010022: ``Measuring, understanding \& reducing respiratory droplet spreading'' and the Netherlands Organisation for Scientific Research (NWO) through the Multiscale Catalytic Energy Conversion (MCEC) research center.

\section*{Declaration of Interests}
The authors report no conflict of interest.

\appendix
\section{Details of the evaporation-condensation analytical model}\label{sec:app}
The evaporation-condensation model used in this paper is developed based on the works of \cite{law1979fuel}, \cite{law1987alcohol}, and \cite{marie2014lagrangian}, and the Fick's law-based multicomponent droplet vapourisation model \citep{tonini2015novel}, with the immiscibility of the HFE-7000--water system taken into consideration. Details of the model are summarised below.

We consider that the gas film surrounding the droplet is spherically symmetric. The evaporation-condensation process is dictated by the vapour diffusion, and the inter-species vapour diffusion is neglected. The droplet is assumed to have a uniform temperature. The system consists of two mass transfer equations and a heat transfer equation of the two-component droplet:
\begin{subequations}
\begin{align}
\frac{\mathrm{d}m_1}{\mathrm{d}t} &= \dot{m}_T\frac{Y_1^s-Y_1^{\infty}\exp\left(\frac{\dot{m}_T}{2\pi\rho^gdD_1}\right)}{1-\exp\left(\frac{\dot{m}_T}{2\pi\rho^gdD_1}\right)}\frac{\Sh_1}{2}\\
\frac{\mathrm{d}m_2}{\mathrm{d}t} &= \dot{m}_T\frac{Y_2^s-Y_2^{\infty}\exp\left(\frac{\dot{m}_T}{2\pi\rho^gdD_2}\right)}{1-\exp\left(\frac{\dot{m}_T}{2\pi\rho^gdD_2}\right)}\frac{\Sh_2}{2}\\
\left(m_1+m_2\right)c_p^d\frac{\mathrm{d}T^d}{\mathrm{d}t} &=\pi d \lambda^g\Nu\frac{\log\left(1+B_T\right)}{B_T}\left(T^{\infty}-T^d\right)+L_1\frac{\mathrm{d}m_1}{\mathrm{d}t}+L_2\frac{\mathrm{d}m_2}{\mathrm{d}t},
\end{align}
\end{subequations}
where $m$ is the liquid mass, $Y$ the vapour mass fraction, $d$ the droplet diameter, $\rho$ the density, $D$ the diffusion coefficient in air, $c_p$ the specific heat capacity, $T$ the temperature, $\lambda$ the thermal conductivity, $L$ the specific latent heat of vaporisation, and $B_T$ the Spalding number of heat transfer (\ref{eq:B_T}). Subscripts 1 and 2 refer to the HFE-7000 and water components, respectively, and $a$ refers to air. Superscript $\infty$ indicates ambient conditions, $d$ refers to the droplet, $s$ refers to droplet surface, and $g$ refers to the vapour gas film around the droplet. Here, $\dot{m}_T$ is the total change rate of the droplet mass without convection correction, which by definition is
\begin{equation}
\dot{m}_T = \frac{\mathrm{d}m_1}{\mathrm{d}t}\frac{2}{\Sh_1}+\frac{\mathrm{d}m_2}{\mathrm{d}t}\frac{2}{\Sh_2}
\end{equation}
and can be found implicitly from the non-linear equation:
\begin{equation}
\frac{Y_1^{\infty}-Y_1^s}{\exp\left(\frac{\dot{m}_T}{2\pi\rho^gdD_1}\right)} + \frac{Y_2^{\infty}-Y_2^s}{\exp\left(\frac{\dot{m}_T}{2\pi\rho^gdD_2}\right)} = 1-Y_1^{\infty}-Y_2^{\infty}.
\end{equation}

The vapour mass fraction at the droplet surface can be expressed as
\begin{subequations}
\begin{align}
Y^s_1 &= \frac{X_1P^s_{\mathrm{sat}1}M_1}{X_1P^s_{\mathrm{sat}1}M_1+X_2P^s_{\mathrm{sat}2}M_2+\left(P^{\infty}-X_1P^s_{\mathrm{sat}1}-X_2P^s_{\mathrm{sat}2}\right) M_a}\\
Y^s_2 &= \frac{X_2P^s_{\mathrm{sat}2}M_2}{X_1P^s_{\mathrm{sat}1}M_1+X_2P^s_{\mathrm{sat}2}M_2+\left(P^{\infty}-X_1P^s_{\mathrm{sat}1}-X_2P^s_{\mathrm{sat}2}\right) M_a}
\end{align}
\end{subequations}
where $P_{\text{sat}}$ is the saturated vapour pressure, $P^{\infty}$ is the ambient pressure, $M$ is molecular weight and $X_1$ and $X_2$ are the surface area fractions of the HFE-7000 and water components, respectively.
\begin{figure}
    \centering
\setlength{\unitlength}{1cm}
\begin{picture}(6,2)
\put(0,0){\includegraphics[scale = 0.3]{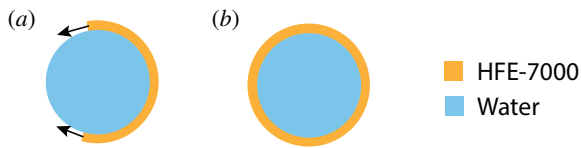}}
\put(-0.5,1.6){(\textit{a})}
\put(2.2,1.6){(\textit{b})}
\end{picture}
    \caption{(\textit{a}) Low-surface-tension component (HFE-7000) wraps around the high-surface-tension water core, driven by the difference in surface tension. (\textit{b}) The wrapping process is complete and HFE-7000 forms a shell around the water core.}
    \label{fig:drop_sketch}
\end{figure}

Unlike typical fuels, which are considered in the model of \cite{sirignano2010fluid}, and also commonly used in the combustion community, water and HFE-7000 have very low solubility in each other \citep[$\lesssim60$ ppmw,][]{NovecTDS}. Therefore, the assumption that liquid in the droplet is fully-mixed is not applicable in our system. HFE-7000 has a very low surface tension of 12.33 mN/m at room temperature \citep{rausch2015density} compared to 72 mN/m for water, so it will wrap around the water droplet core, driven by the surface tension, as shown in figure \ref{fig:drop_sketch}(\textit{a}). For a water core with a diameter of 2~$\upmu$m, the time required for the HFE-7000 film to completely wrap around it is estimated to be $30$~ns \citep{koldeweij2019marangoni}, which is much short than the droplet lifetime. Therefore, we consider a `shell' model of the evaporating droplet as depicted in figure \ref{fig:drop_sketch}(\textit{b}). In this scenario, the surface area fractions are $X_1 = 1$ and $X_2 = 0$. When HFE-7000 is fully evaporated, the water core will be fully exposed and the surface area fractions therefore become $X_1 = 0$ and $X_2 = 1$.

The reference mass fractions and temperature in the gas film are estimated using the so-called 1/3 law \citep{hubbard1975droplet}:
\begin{subequations}
\begin{align}
Y_1^r = \frac{2}{3}Y_1^s+\frac{1}{3}Y_1^{\infty}\\
Y_2^r = \frac{2}{3}Y_2^s+\frac{1}{3}Y_2^{\infty}\\
T^r = \frac{2}{3}T^s+\frac{1}{3}T^{\infty}
\end{align}
\end{subequations}

The gas film density is computed from a weighted harmonic mean of the compositions:
\begin{equation}
\rho^g = \frac{1}{\frac{Y^r_1}{\rho_1}+ \frac{Y^r_2}{\rho_2}+\frac{1-Y^r_1-Y^r_2}{\rho_a}}.
\end{equation}

The Spalding number of mass transfer is a non-dimensional thermodynamics parameter measuring the ratio of drive towards vaporisation as compared to resistance to vaporisation \citep{abramzon1989droplet}:
\begin{equation}
B_M = \frac{Y_1^s-Y_1^{\infty}+Y_2^s-Y_2^{\infty}}{1-Y_1^s-Y_2^s},
\end{equation}
and the Spalding number of heat transfer $B_T$ can be linked to $B_M$ through \citep{marie2014lagrangian}:
\begin{equation}
\log\left(1+B_T\right) = \frac{c_p^g D^g\rho^g}{\lambda^g}\log\left(1+B_{M}\right).
\label{eq:B_T}
\end{equation}

Finally, the effect of turbulent convection is incorporated using the so-called `film theory', which assumes that the resistance to heat or mass exchange between a surface and a gas flow may be modelled by introducing the concept of a surrounding gas film \citep{frank1969diffusion, abramzon1989droplet}. The Sherwood and Nusselt numbers can be expressed as \citep{abramzon1989droplet}:
\begin{subequations}
\begin{align}
\Sh_1&= 2+\frac{\Sh_{01}-2}{F(B_{M})},\label{eq:Sh1}\\
\Sh_2&= 2+\frac{\Sh_{02}-2}{F(B_{M})},\\
\Nu &= 2+\frac{\Nu_{0}-2}{F(B_{T})},
\end{align}
\end{subequations}
where $\Sh_0$ is the Sherwood number and $\Nu_0$ the Nusselt number for a non-vapourising sphere \citep{abramzon1989droplet}:
\begin{subequations}
\begin{align}
\Sh_{01} &= 1+\left(1+\Rey_d \Sc_1^g\right)^{1/3}\max\left(1,\Rey_d^{0.077}\right),\\
\Sh_{02} &= 1+\left(1+\Rey_d \Sc_2^g\right)^{1/3}\max\left(1,\Rey_d^{0.077}\right),\\
\Nu_{0} &= 1+\left(1+\Rey_d \Pr^g\right)^{1/3}\max\left(1,\Rey_d^{0.077}\right),\\
&\text{for } \Rey_d\leq5,
\end{align}
with
\end{subequations} 
\begin{equation}
\Sc_1^g = \frac{\mu^g}{D_1\rho^g},\quad
\Sc_2^g = \frac{\mu^g}{D_2\rho^g},\quad
\Pr^g = \frac{\mu^g c_p^g}{\lambda^g},
\end{equation}
where $\mu^g$ is the dynamic viscosity of the gas film around the droplet. The term $F(B)$ represents the relative change of the gas film thickness due to the Stefan flow around the droplet, which is given by \citep{abramzon1989droplet}:
\begin{equation}
\begin{split}
F(B) &= (1+B)^{0.7}\frac{\log(1+B)}{B},\quad B = B_M, B_T,\\
&\text{for } 0\leq B_T, \, B_M\leq20, \, 1\leq\Pr,\,\Sc\leq3.
\end{split}
\end{equation}

\bibliographystyle{jfm}
\bibliography{arxiv.bib}

\begin{thebibliography}{25}
\expandafter\ifx\csname natexlab\endcsname\relax\def\natexlab#1{#1}\fi
\def\au#1{#1} \def\ed#1{#1} \def\yr#1{#1}\def\at#1{#1}\def\jt#1{\textit{#1}}
  \def\bt#1{#1}\def\bvol#1{\textbf{#1}} \def\vol#1{#1} \def\pg#1{#1}
  \def\publ#1{#1}\def\arxiv#1{#1}\def\org#1{#1}\def\st#1{\textit{#1}}

\bibitem[3M(2021)]{NovecTDS}
{\sc \au{3M}} \yr{2021} {N}ovec\texttrademark { }7000 {E}ngineered {F}luid
  technical data sheet.

\bibitem[Abramzon \& Sirignano(1989)]{abramzon1989droplet}
{\sc \au{Abramzon, B.} \& \au{Sirignano, W.~A.}} \yr{1989}  \at{Droplet
  vaporization model for spray combustion calculations}.  \jt{Int. J. Heat Mass
  Transf.}  \bvol{32}~(9),  \pg{1605--1618}.

\bibitem[{Al Zaitone}(2018)]{ALZAITONE2018164}
{\sc \au{{Al Zaitone}, B.}} \yr{2018}  \at{Oblate spheroidal droplet
  evaporation in an acoustic levitator}.  \jt{Int. J. Heat Mass Transf.}
  \bvol{126},  \pg{164--172}.

\bibitem[Birouk \& G{\"o}kalp(2006)]{birouk2006current}
{\sc \au{Birouk, M.} \& \au{G{\"o}kalp, I.}} \yr{2006}  \at{Current status of
  droplet evaporation in turbulent flows}.  \jt{Prog. Energy Combust. Sci.}
  \bvol{32}~(4),  \pg{408--423}.

\bibitem[Da\"if {\em et~al.\/}(1998)Da\"if, Bouaziz, Chesneau \&
  Ch\'erif]{daif1998comparison}
{\sc \au{Da\"if, A.}, \au{Bouaziz, M.}, \au{Chesneau, X.} \& \au{Ch\'erif,
  A.~A.}} \yr{1998}  \at{Comparison of multicomponent fuel droplet vaporization
  experiments in forced convection with the {S}irignano model}.  \jt{Exp.
  Therm. Fluid Sci.}  \bvol{18}~(4),  \pg{282--290}.

\bibitem[Dalla~Barba {\em et~al.\/}(2021)Dalla~Barba, Wang \&
  Picano]{dalla2021revisiting}
{\sc \au{Dalla~Barba, F.}, \au{Wang, J.} \& \au{Picano, F.}} \yr{2021}
  \at{Revisiting {D}$^2$-law for the evaporation of dilute droplets}.
  \jt{Phys. Fluids}  \bvol{33}~(5),  \pg{051701}.

\bibitem[Dodd {\em et~al.\/}(2021)Dodd, Mohaddes, Ferrante \&
  Ihme]{dodd2021analysis}
{\sc \au{Dodd, M.~S.}, \au{Mohaddes, D.}, \au{Ferrante, A.} \& \au{Ihme, M.}}
  \yr{2021}  \at{Analysis of droplet evaporation in isotropic turbulence
  through droplet-resolved {DNS}}.  \jt{Int. J. Heat Mass Transf.}  \bvol{172},
   \pg{121157}.

\bibitem[Frank-Kamenetski(1969)]{frank1969diffusion}
{\sc \au{Frank-Kamenetski, D.~A.}} \yr{1969} {\em Diffusion and heat transfer
  in chemical kinetics\/}.  \publ{Plenum Press}.

\bibitem[Hubbard {\em et~al.\/}(1975)Hubbard, Denny \&
  Mills]{hubbard1975droplet}
{\sc \au{Hubbard, G.~L.}, \au{Denny, V.~E.} \& \au{Mills, A.~F.}} \yr{1975}
  \at{Droplet evaporation: effects of transients and variable properties}.
  \jt{Int. J. Heat Mass Transf.}  \bvol{18}~(9),  \pg{1003--1008}.

\bibitem[Koldeweij {\em et~al.\/}(2019)Koldeweij, Van~Capelleveen, Lohse \&
  Visser]{koldeweij2019marangoni}
{\sc \au{Koldeweij, R. B.~J.}, \au{Van~Capelleveen, B.~F.}, \au{Lohse, D.} \&
  \au{Visser, C.~W.}} \yr{2019}  \at{Marangoni-driven spreading of miscible
  liquids in the binary pendant drop geometry}.  \jt{Soft Matter}
  \bvol{15}~(42),  \pg{8525--8531}.

\bibitem[Langmuir(1918)]{langmuir1918evaporation}
{\sc \au{Langmuir, I.}} \yr{1918}  \at{The evaporation of small spheres}.
  \jt{Phys. Rev.}  \bvol{12}~(5),  \pg{368}.

\bibitem[Law \& Binark(1979)]{law1979fuel}
{\sc \au{Law, C.~K.} \& \au{Binark, M.}} \yr{1979}  \at{Fuel spray vaporization
  in humid environment}.  \jt{Int. J. Heat Mass Transf.}  \bvol{22}~(7),
  \pg{1009--1020}.

\bibitem[Law {\em et~al.\/}(1987)Law, Xiong \& Wang]{law1987alcohol}
{\sc \au{Law, C.~K.}, \au{Xiong, T.~Y.} \& \au{Wang, C.}} \yr{1987}
  \at{Alcohol droplet vaporization in humid air}.  \jt{Int. J. Heat Mass
  Transf.}  \bvol{30}~(7),  \pg{1435--1443}.

\bibitem[Mari{\'e} {\em et~al.\/}(2014)Mari{\'e}, Grosjean, M{\'e}{\`e}s,
  Seifi, Fournier, Barbier \& Lance]{marie2014lagrangian}
{\sc \au{Mari{\'e}, J.-L.}, \au{Grosjean, Na.}, \au{M{\'e}{\`e}s, L.},
  \au{Seifi, M.}, \au{Fournier, C.}, \au{Barbier, B.} \& \au{Lance, M.}}
  \yr{2014}  \at{Lagrangian measurements of the fast evaporation of falling
  diethyl ether droplets using in-line digital holography and a high-speed
  camera}.  \jt{Exp. Fluids}  \bvol{55}~(4),  \pg{1--13}.

\bibitem[M{\'e}{\`e}s {\em et~al.\/}(2020)M{\'e}{\`e}s, Grosjean, Mari{\'e} \&
  Fournier]{mees2020statistical}
{\sc \au{M{\'e}{\`e}s, L.}, \au{Grosjean, N.}, \au{Mari{\'e}, J.-L.} \&
  \au{Fournier, C.}} \yr{2020}  \at{Statistical {L}agrangian evaporation rate
  of droplets released in a homogeneous quasi-isotropic turbulence}.  \jt{Phys.
  Rev. Fluids}  \bvol{5}~(11),  \pg{113602}.

\bibitem[Promvongsa {\em et~al.\/}(2017)Promvongsa, Vallikul, Fungtammasan,
  Garo, Grehan \& Saengkaew]{promvongsa2017multicomponent}
{\sc \au{Promvongsa, J.}, \au{Vallikul, P.}, \au{Fungtammasan, B.}, \au{Garo,
  A.}, \au{Grehan, G.} \& \au{Saengkaew, S.}} \yr{2017}  \at{Multicomponent
  fuel droplet evaporation using 1{D} global rainbow technique}.  \jt{Proc.
  Combust. Inst.}  \bvol{36}~(2),  \pg{2401--2408}.

\bibitem[Ra \& Reitz(2009)]{ra2009vaporization}
{\sc \au{Ra, Y.} \& \au{Reitz, R.~D.}} \yr{2009}  \at{A vaporization model for
  discrete multi-component fuel sprays}.  \jt{Int. J. Multiph. Flow}
  \bvol{35}~(2),  \pg{101--117}.

\bibitem[Rausch {\em et~al.\/}(2015)Rausch, Kretschmer, Will, Leipertz \&
  Fr{\"o}ba]{rausch2015density}
{\sc \au{Rausch, M.~H.}, \au{Kretschmer, L.}, \au{Will, S.}, \au{Leipertz, A.}
  \& \au{Fr{\"o}ba, A.~P.}} \yr{2015}  \at{Density, surface tension, and
  kinematic viscosity of hydrofluoroethers {HFE}-7000, {HFE}-7100, {HFE}-7200,
  {HFE}-7300, and {HFE}-7500}.  \jt{J. Chem. Eng. Data}  \bvol{60}~(12),
  \pg{3759--3765}.

\bibitem[Sahu {\em et~al.\/}(2018)Sahu, Hardalupas \&
  Taylor]{sahu2018interaction}
{\sc \au{Sahu, S.}, \au{Hardalupas, Y.} \& \au{Taylor, A. M. K.~P.}} \yr{2018}
  \at{Interaction of droplet dispersion and evaporation in a polydispersed
  spray}.  \jt{J. Fluid Mech.}  \bvol{846},  \pg{37--81}.

\bibitem[Sirignano(2010)]{sirignano2010fluid}
{\sc \au{Sirignano, W.~A.}} \yr{2010} {\em Fluid dynamics and transport of
  droplets and sprays\/}.  \publ{Cambridge University Press}.

\bibitem[Tonini \& Cossali(2015)]{tonini2015novel}
{\sc \au{Tonini, S.} \& \au{Cossali, G.~E.}} \yr{2015}  \at{A novel formulation
  of multi-component drop evaporation models for spray applications}.  \jt{Int.
  J. Therm. Sci.}  \bvol{89},  \pg{245--253}.

\bibitem[Tonini \& Cossali(2019)]{tonini2019analytical}
{\sc \au{Tonini, S.} \& \au{Cossali, G.~E.}} \yr{2019}  \at{An analytical
  approach to model heating and evaporation of multicomponent ellipsoidal
  drops}.  \jt{Heat Mass Transf.}  \bvol{55}~(5),  \pg{1257--1269}.

\bibitem[Villermaux {\em et~al.\/}(2017)Villermaux, Moutte, Amielh \&
  Meunier]{villermaux2017fine}
{\sc \au{Villermaux, E.}, \au{Moutte, A.}, \au{Amielh, M.} \& \au{Meunier, P.}}
  \yr{2017}  \at{Fine structure of the vapor field in evaporating dense
  sprays}.  \jt{Phys. Rev. Fluids}  \bvol{2}~(7),  \pg{074501}.

\bibitem[Wang {\em et~al.\/}(2021)Wang, Dalla~Barba \& Picano]{wang2021direct}
{\sc \au{Wang, J.}, \au{Dalla~Barba, F.} \& \au{Picano, F.}} \yr{2021}
  \at{Direct numerical simulation of an evaporating turbulent diluted jet-spray
  at moderate {R}eynolds number}.  \jt{Int. J. Multiph. Flow}  \bvol{137},
  \pg{103567}.

\bibitem[Yarin {\em et~al.\/}(1999)Yarin, Brenn, Kastner, Rensink \&
  Tropea]{yarin1999evaporation}
{\sc \au{Yarin, A.~L.}, \au{Brenn, G.}, \au{Kastner, O.}, \au{Rensink, D.} \&
  \au{Tropea, C.}} \yr{1999}  \at{Evaporation of acoustically levitated
  droplets}.  \jt{J. Fluid Mech.}  \bvol{399},  \pg{151^^e2^^80^^93204}.

\bibitem[Tonini \& Cossali (2016)]{tonini2016one}
{\sc \au{Tonini, S.} \& \au{Cossali, G.~E.}} \yr{2016}
\at{One-dimensional analytical approach to modelling evaporation and heating of deformed drops}.
\jt{Int. J. Heat Mass Transf.}  \bvol{97}, \pg{301--307}.

\end{thebibliography}
\end{document}